\font\mybb=msbm10 at 12pt
\def\bb#1{\hbox{\mybb#1}}

\def\R {\bb{R}}
\def\E {\bb{E}}
\tolerance=10000
\input phyzzx

\def\unit{\hbox to 3.3pt{\hskip1.3pt \vrule height 7pt width .4pt \hskip.7pt
\vrule height 7.85pt width .4pt \kern-2.4pt
\hrulefill \kern-3pt
\raise 4pt\hbox{\char'40}}}


\REF\a{H.L{\" u}, C.N.Pope, E.Sezgin and K.S.Stelle, {\sl Stainless
super p-branes}, Nucl. Phys. {\bf B456} (1995) 669, hep-th/9508042.}
\REF\b{L.Romans, {\sl Massive N=2A supergravity in ten dimensions},
Phys. Lett. {\bf B169} (1986) 374.}
\REF\c{P.K.Townsend and P van Nieuwenhuizen, {\sl Gauged
seven-dimensional supergravity}, Phys. Lett. {\bf B125} (1983) 41.}
\REF\d{L.Romans, {\sl The F(4) gauged supergravity in six dimensions},
Nucl. Phys. {\bf B269} (1986) 691.}
\REF\e{D.Z.Freedman and J.H.Schwarz, {\sl N=4 Supergravity theory with
local SU(2)$\times$SU(2) invariance}, Nucl. Phys. {\bf B137} (1978) 333.}
\REF\f{D.Z.Freedman and G.W.Gibbons, {\sl Electrovac ground state in
gauged SU(2)$\times$SU(2) supergravity}, Nucl. Phys, {\bf B233} (1984)
24.}
\REF\g{E.Bergshoeff, M.de Roo, M.B.Green, G.Papadopoulos and
P.K.Townsend, {\sl Duality of type II 7-branes and 8-branes},
Nucl. Phys. {\bf B470} (1996) 113, hep-th/9601150.}
\REF\h{H.L{\" u}, C.N.Pope, E.Sezgin and K.S.Stelle, {\sl Dilatonic 
p-brane solitons}, Phys. Lett. {\bf B371} (1996) 46, hep-th/9511203.}
\REF\i{H.L{\" u}, C.N.Pope and P.K.Townsend, {\sl Domain walls from
Anti-de Sitter spacetime}, Phys. Lett. {\bf B391} (1997) 39,
hep-th/9607164.}
\REF\j{K.S.Stelle, {\sl BPS branes in supergravity}, hep-th/9803116.}
\REF\k{P.M.Cowdall, H.L{\" u}, C.N.Pope, K.S.Stelle and P.K.Townsend,
{\sl Domain walls in massive supergravities}, Nucl. Phys. {\bf B486}
(1997) 49, hep-th/9608173.}
\REF\l{A.H. Chamseddine and M.S. Volkov, {\sl Non-abelian solitons in 
N=4 gauged supergravity and leading order string theory}, hep-th/9711181.}
\REF\m{F.Giani, M.Pernici and P van Nieuwenhuizen, {\sl Gauged N=4 
supergravity in six dimensions}, Phys. Rev. {\bf D30} (1984) 1680.}
\REF\n{F. Quevedo, {\sl Electrovac compactification of gauged d=7 
supergravity}, Phys. Lett. {\bf B148} (1984) 301.}
\REF\o{E.Cremmer, B.Julia and J.Scherk, {\sl Supergravity theory in
eleven dimensions}, Phys. Lett. {\bf B76} (1978) 409.}
\REF\p{I.Antoniadis, C.Bachas and A.Sagnotti, {\sl Gauged supergravity
vacua in string theory}, Phys. Lett. {\bf B235} (1990) 255.}
\REF\q{G.W.Gibbons, D.Z.Freedman and P.C.West, {\sl Ten into four
won't go}, Phys. Lett. {\bf B124} (1983) 491.}
\REF\r{M.Pernici, K.Pilch and P.Van Nieuwenhuizen, {\sl Gauged
maximally extended supergravity in seven dimensions}, Phys. Lett. {\bf
B143} (1984) 103.}
\REF\s{L.J.Romans, {\sl Gauged N=4 supergravities in five dimensions
and their magnetovac backgrounds}, Nucl. Phys. {\bf B267} (1986) 433.}
\REF\t{P.K. Townsend, {\sl Anti-de Sitter supergravities}, in 
Quantum Field Theory and Quantum Statistics, eds. I.A. Batalin,
C.J. Isham, and G.A. Vilkovisky (Adam Hilger 1987) pp. 299-308.}
\REF\u{A.Salam and E.Sezgin, {\sl Maximal extended supergravity in
seven dimensions}, Phys. Lett. {\bf B118} (1982) 359.}


\Pubnum{ \vbox{ \hbox{DAMTP-R/97/51}}}
\pubtype{hep-th/9710214 }
\date{}

\titlepage

\title {\bf Supersymmetric Electrovacs In Gauged Supergravities}

\author{P.M.Cowdall\foot{p.m.cowdall@damtp.cam.ac.uk}}
\address{DAMTP, Silver St.,
\break
Cambridge CB3 9EW, U.K.}

\abstract{We show that the D=6 SU(2) gauged supergravity of van
Nieuwenhuizen et al, obtained by dimensional reduction of the D=7
topologically massive gauged supergravity and previously thought not
to be dimensionally reducible, can be further reduced to five and four
dimensions. On reduction to D=4 one recovers the special case of 
the SU(2)$\times$SU(2) gauged
supergravity of Freedman and Schwarz for which one of the SU(2)
coupling constants vanishes. Previously known supersymmetric
electrovacs of this model then imply new ground states in 7-D. We
construct a supersymmetric electrovac solution of N=2 SU(2) gauged
supergravity in 7-D. We also investigate the domain wall
solutions of these theories and show they preserve a half of the
supersymmetry.} 

\endpage


{\bf \chapter{\bf Introduction}}

There are many supergravity theories with a scalar potential without
critical points and hence no obvious ground state. In many simple
cases the scalar potential takes the form

$$
V(\phi)\sim e^{-a\phi}
\eqn\zeroa
$$
where $a$ is a constant which is related to a very useful quantity $\Delta$,
by the equation
$$
a^2 = \Delta + {2(D-1)\over(D-2)}
\eqn\zerob
$$
where D is the spacetime dimension.\foot {we assume the dilaton is
canonically normalised} It is convenient to reexpress the parameter
$a$ in terms of $\Delta$ because
$\Delta$ remains unchanged after reduction on $S^1$ [\a]. Moreover,
$\Delta$ also allows us to distinguish between massive and gauged
supergravities with potentials of the above form. Massive
supergravities are theories that can contain topological mass terms
as well as explicit mass terms for some of the antisymmetric tensor
potentials in the model. The value of $\Delta$ for these theories is
4. A classic example of such a supergravity is the D=10 massive IIA
supergravity [\b], in which a second rank antisymmetric tensor
acquires a mass in a Higgs type mechanism. Gauged supergravities on
the other hand, are distinct from massive supergravities in that the
automorphism group (or one of its subgroups) of the supersymmetry
algebra is gauged, the vector fields of the supergravity multiplet
playing the role of the gauge fields. The subclass of gauged
supergravities with potentials like (1.1) have $\Delta$ negative in
all known cases. Well known examples are the D=7 N=2 SU(2) gauged
supergravity [\c], D=6 N=4 SU(2) gauged supergravity [\d] and the D=4
N=4 SU(2)$\times$SU(2) model of Freedman and Schwarz (FS) [\e]. The
value of $\Delta$ for all these SU(2) gauged supergravities is
-2. As $\Delta$ is unchanged under both Kaluza-Klein (KK) and
Scherk Schwarz reduction on $S^1$, one might suspect that these
theories are related. This is in fact true but until recently 
it was believed these
supergravities couldn't be dimensionally reduced. The reason why it
is now believed that they can be will be explained later.

At this point we note that there also exist supergravities
whose potentials are sums of functions of the type given in
(1.1). Supergravities which are both gauged and possessing topological
mass terms are examples of these types of theories.

There are two important questions to be asked of supergravities
with potentials as in (1.1). The first concerns the nature of the
ground state of these theories and the second concerns whether these
theories are consistent truncations of higher dimensional
theories. Due to the lack of a Minkowski vacuum preserving all of the
supersymmetry, it is natural to look for solutions preserving some
fraction of the supersymmetry. These supersymmetric solutions would of
course be stable and therefore presumably important to understanding
whether these theories make sense. The FS model and it's
SU(2)$\times$U(1)$^3$ `half gauged' version, obtained by setting one
of the SU(2) coupling constants to zero, are known to have
supersymmetric electrovac solutions [\f]. These supersymmetric
electrovac solutions have the property that the dilaton is zero and a
two form gauge field strength is covariantly constant. The D=10
massive type IIA supergravity and the D=7 N=2 SU(2) gauged
supergravity possess domain wall ((D-2) brane) solutions preserving a
half of the supersymmetry [\g,\h,\i]. These solutions differ from the
much studied p-brane solutions [\j] \foot {p $\not=$ D-2}
 in that the dilaton is the only
non-constant non-vanishing field and so also have a quite different
geometry to the supersymmetric electrovacs. A further question of
interest then is whether the FS model has supersymmetric domain wall
solutions and whether the D=7 N=2 SU(2) supergravity has
supersymmetric electrovac solutions. This will be answered by
looking at the question of dimensional reduction.

 Because of the form of $V(\phi)$, these supergravities admit no
$S^{1}\times M_{D-1}$ direct product vacuum solutions so it might seem
that dimensional reduction is not possible. However, it was argued in
[\i] that one can always perform a consistent dimensional reduction,
regardless of the solution space, by simply implementing the standard
KK ansatz on the fields of the higher dimensional
theory (substitution of the standard KK ansatz for the fields into the
lagrangian gives a lower dimensional lagrangian whose field equations
are the same as those obtained by direct substitution of the KK ansatz
into the higher dimensional field equations). However, one is not
guaranteed solutions of the lower dimensional theory unless, in the
higher dimension, there exists a solution with a U(1) isometry, in
which case it will map into a solution of the lower dimensional
theory. The dimensional reduction of massive supergravities was
studied in [\k]. In this paper we focus on the gauged supergravities
with SU(2) gauge groups. Now the FS model and the D=7 SU(2) gauged
model have no known higher dimensional origin.\foot {At the time of
writing a paper has appeared [\l] showing D=10 type I
supergravity on $S^3{\times}S^3$ yields the Freedman Schwarz
model. The implications of this result will be explored in future
publications.} It was believed these supergravities couldn't be
dimensionally reduced, for reasons given above, and hence were
unrelated. However van Nieuwenhuizen et al [\m] argued that with the
inclusion of a topological mass term, the scalar potential of N=2
SU(2) gauged 7-D supergravity depends on two parameters and does
posses a stable minimum, so therefore can be compactified to 6-D. They
showed that if the parameter in front of the topological mass term is
non-zero, the resulting 6-D theory is irreducible. They also showed
that if this parameter is allowed to tend to zero in 6-D, the theory
describes the reducible coupling of an SU(2) gauged pure N=4
supergravity multiplet to an N=4 vector supermultiplet. This D=6 N=4
SU(2) gauged supergravity was a special case of the massive gauged
supergravities presented in [\d].\foot{Following [\i], we note that
the D=6 N=4 SU(2) gauged supergravity obtained in [\m] can also be
obtained by dimensional reduction of D=7 N=2 SU(2) gauged supergravity
without a topological mass term.}  

One of the purposes of this paper
therefore is to continue this reduction to four dimensions, making
contact with the SU(2)$\times$SU(2) gauged model of Freedman and
Schwarz. Given we know the D=7 SU(2) gauged supergravity to have a
domain wall solution with an R-isometry \foot {R-isometries are as
good as U(1) isometries for the purposes of this argument.}, the
reduction will be consistent and we are guaranteed solutions in the
lower dimension. We will show that D=6 N=4 SU(2) gauged supergravity
can be reduced on T$^2$ to yield, after appropriate truncations, a
version of the SU(2)$\times$SU(2) FS model for which one of the SU(2)
gauge coupling constants vanishes. This relation between the D=7 and
D=4 theories has implications for their solutions. In particular we
must be able to lift the supersymmetric electrovac of the `half
gauged' FS model from 4-D to 7-D. We use this fact to construct a
previously unknown and supersymmetric electrovac in 7-D. We are also able
to double dimensionally reduce the 7-D supersymmetric domain wall 
to give a supersymmetric domain wall of the FS model.

The layout of the paper is as follows. In section 2 we dimensionally
reduce the bosonic sector of the 6-D supergravity of [\m] to
 5-D giving details of how to
truncate out the bosonic fields of a vector supermultiplet.  In
section 3 we continue this reduction to 4-D in less detail and make
contact with the FS model. In section 4 we solve the Killing spinor
equations of SU(2) gauged 7-D supergravity to obtain a supersymmetric
electrovac preserving a half of the supersymmetry. In section 5 we
double dimensionally reduce the 1/2 supersymmetric domain wall
solution of 7-D gauged supergravity to obtain a domain wall in 4-D
(which could have been found directly but this method demonstrates the
consistency of the reduction). In section 6 we show that this domain
wall preserves a half of the supersymmetry.


{\bf \chapter{\bf Dimensional Reduction Of D=6 Gauged Supergravity }}

We now turn to the reduction of the N=4 SU(2) gauged supergravity in
6-D. We do not consider the reduction of the fermionic sector of the
lagrangian or the supersymmetry transformations. However, it can be
argued that under dimensional reduction (i.e. assuming all fields are
independent of a particular coordinate) any symmetry of the higher
dimensional lagrangian remains a symmetry of the lower dimensional
lagrangian (this is not always true at the level of solutions
though). Hence supersymmetry will be preserved when we reduce to five
and four dimensions and our bosonic lagrangians are guaranteed
supersymmetric extensions. We therefore consider the reduction of the
bosonic sector only, which is [\m] \foot{We use the mostly plus metric
convention.}
$$
\eqalign{
  L = {\hat e}\big\{ {R_6} -{1\over 2}(\partial_{\mu}{\hat {\phi}})^2 -{1\over
    4}e^{-{{\hat{\phi}}\over \sqrt{2}}} [ ({\hat F_{\mu\nu i}}^{\; \;
    \; \; \;j})^2 +
  ({\hat G_{\mu\nu}})^2 ]  -{1\over12}e^{2{\hat{\phi}}\over\sqrt{2}}
  ({\hat F_{\mu\nu\rho}})^2 + 4\alpha^{2} e^{{\hat{\phi}}\over\sqrt2} \big\}\cr   
  -{1\over24} \epsilon^{\mu\nu\rho\sigma\lambda\delta}\hat F_{\mu\nu\rho}[
  \hat G_{\sigma\lambda}\hat B_{\delta} + Tr(\hat
    F_{\sigma\lambda}\hat A_{\delta}
  -{2i\alpha\over3}\hat A_{\sigma}\hat A_{\lambda}\hat A_{\delta}) ] } 
\eqn\onea  
$$
where $\hat F_{\mu\nu\rho} = 3\partial_{[\mu} {\hat A_{\nu\rho ]}}$ ,
$ \hat G_{\mu\nu} = 2\partial_{[\mu}\hat B_{\nu ]} $ and ${{\hat
F}_{\mu\nu i}}^{\; \; \; \; \; j} = 2(\partial_{[\mu}^{\ } {\hat
A_{\nu ] i}}^{\; \; \; \; j} + i\alpha{\hat A_{[\mu |i|}}^{\; \; \; \;
\; \; k}{\hat A_{\nu ] k}}^{\; \; \;
\; j})$. ${{\hat F}_{\mu\nu i}}^{\; \; \; \; \; j}$ is a 2$\times$2
symmetric matrix with i,j = 1,2. To perform the reduction to five dimensions, the ansatz for the fields are :
$$
{\hat e}^{\hat a}_{\hat m} = \pmatrix{ e^{\sigma\over2\sqrt6}e^{a}_{m} &0\cr
e^{-{3\sigma\over2\sqrt6}}\tilde A_{m} & e^{-{3\sigma\over2\sqrt6}}\cr } 
\eqn\oneb
$$ 
Where hats refer to D=6.
$\hat a =(a,\underline z)$ are local Lorenz indices.
$\hat m =(m,z)$ are world indices.
$$ 
d{\hat S_{6}}^{2}=e^{\sigma\over\sqrt6}dS_{5}^{2}+e^{-{3\sigma\over\sqrt6}} 
(dz+\tilde A)^{2}
\eqn\onec
$$
Where $ \tilde A = \tilde A_{m}dx^{m}$ and $  f_{2} =d\tilde A.$
$$
\hat B_{1} = B_{1}^{(1)}  +  \rho dz
\eqn\oned
$$
$$
\hat A_{2} = A_{2} + C_{1} dz
\eqn\onee
$$
$$
{{{\hat A}}_{1 i}}^{\; \; \; j} = {A_{1 i}}^{j} + {A_{i}}^{j} dz
\eqn\onef
$$
$$
\hat \phi (x^m,z) = \phi (x^m)
\eqn\oneg
$$

Since we are reducing a non-maximal
supergravity theory, in 5-D one necessarily obtains
another non-maximal supergravity coupled to other supermultiplets, in
this case a vector supermultiplet. We are interested only in the
supergravity multiplet so we need to truncate out the vector
supermultiplet. For the truncation to be consistent all one requires
is for the full field equations to allow the fields to be truncated to
be set to zero. For the truncation to preserve supersymmetry we
require the variation of the truncated fermions to vanish (i.e. to be
at least linear in the fields that are set to zero in the
truncation). We have not explicitly shown this but we add that the
final results suggest that supersymmetry is in fact preserved.

An N=4 vector supermultiplet in 5-D contains one vector 
and five scalar bosonic degrees of freedom. Now in [\m], it was shown
that in order
to consistently truncate out the N=4 vector supermultiplet it is
necessary to linearly combine the dualisation of the antisymmetric 
potential $A_{\mu\nu\rho}$ with the KK vector and it is
necessary to linearly combine the dilaton (from the dilaton in 7-D) 
with the KK scalar. This suggests similar tasks must be
undertaken 
in 5-D and 4-D to consistently truncate out the vector supermultiplets.
Proceeding in the same spirit we retain the scalars $\sigma$ and
$\phi$ which we will later linearly combine using an SO(2) rotation.
We need to set four of the scalars to zero, we choose these to be
${A_{i}}^{j}$ and $\rho$. The resulting action in 5-D is
$$
\eqalign{
S_{5} = \int d^{5}x e\big\{ R_{5} - {1\over2}{|d\sigma|}^{2} -{1\over4}
e^{-{4\sigma\over\sqrt{6}}} |f_{2}|^{2} -{1\over2}{|d\phi|}^{2}-{1\over4} 
e^{-{\phi\over\sqrt{2}} -{\sigma\over\sqrt{6}}} [ |F_{2}|^{2} +
|G_{2}^{(1)}|^{2} ] \cr
 -{1\over12}e^{{2\phi\over\sqrt{2}} -
{2\sigma\over\sqrt{6}}} |F_{3}|^{2} -{1\over4}e^{{2\phi\over\sqrt{2}} +
{2\sigma\over\sqrt{6}}} |C_{2}|^{2} + 4{\alpha}^{2}e^{{\phi\over\sqrt{2}}
+ {\sigma\over\sqrt{6}}} \big\} \cr
+ {1\over2}\int C_{2}[G_{2}^{(1)} B_{1}^{(1)} +
Tr( F_{2} A_{1} - {i\alpha\over3}A_{1}A_{1}A_{1} )] }
\eqn\oneh
$$
where $F_{2}=dA_{1} +i\alpha A_{1}A_{1}$, $G_{2}^{(1)}=dB_{1}^{(1)}$, 
$F_{3}=dA_{2}-dC_{1}{\tilde A}$, $C_{2}=dC_{1}$ and wedge products are
implied in the Chern-Simons term.

We still need to truncate a vector and a scalar. As a first step
towards truncating a
vector we dualise $A_{\mu\nu}$ to a vector
${A_{\mu}^{\prime}}$ and linearly combine with the KK
vector. The relevant part of the action is 
$$
\eqalign{
{S_{5}^{\prime}} = -{1\over12}\int d^{5}x
ee^{{2\phi\over\sqrt{2}}-{2\sigma\over\sqrt{6}}}
(F_{\mu\nu\rho}F^{\mu\nu\rho}) = \int d^{5}x L }
\eqn\onei
$$
where $F_{\mu\nu\rho} =
3({\partial_{[ \protect \mu} A_{\nu\rho  ]}}-2({\partial_{[ \protect \mu} C_{\nu }){\tilde 
A_{\rho ]}}}).$

We replace $3{\partial_{[ \protect \mu}A_{\nu\rho ]}}$ by a
independent field $a_{\mu\nu\rho}$.  Let $X_{\mu\nu\rho} =
-6\partial_{[\protect \mu}C_{\nu }{\tilde A_{\rho ]\protect }}$ and
add to the Lagrangian ,
$$
\Delta L = a{\epsilon^{\mu\nu\rho\sigma\gamma}}a_{\mu\nu\rho}
{F^{\prime} _{\sigma\gamma}}
\eqn\onej
$$
where $F_{\sigma\gamma}^{\prime} = 2{\partial^{\ }_{[\protect \sigma}}
A_{\gamma ]\protect }^{\prime}$ and $a$ is a constant.
$$
L + {\Delta L} = -{e\over12}e^{{2\phi\over\sqrt{2}}-{2\sigma\over\sqrt{6}}}
[ a_{\mu\nu\rho}+X_{\mu\nu\rho}][a^{\mu\nu\rho}+X^{\mu\nu\rho}] +
a{\epsilon^{\mu\nu\rho\sigma\gamma}}a_{\mu\nu\rho}{F^{\prime} _{\sigma\gamma}}
\eqn\onek
$$
Variation w.r.t. $A_{\mu}^{\prime}$ gives $a_{\mu\nu\rho} =
3{\partial_{[\mu} A_{\nu\rho ]}}$, and the original Lagrangian, $L$,
can be recovered up to a total derivative after substitution.

\noindent 
Variation w.r.t $a_{\mu\nu\rho}$ leads to
$$
a_{\mu\nu\rho}=6aee^{-{2\phi\over\sqrt{2}}+{2\sigma\over\sqrt{6}}}
{\epsilon_{\mu\nu\rho\sigma\gamma}}{F^{\prime\sigma\gamma}}-X_{\mu\nu\rho}
\eqn\onel
$$
which upon substitution back in the intermediate Lagrangian,
 $L+{\Delta L}$,
 and choosing the constant $a={+{1\over12}}$
for correct normalisation, we obtain the dual Lagrangian $L_D$,
$$
L_D = 
-{e\over4}e^{-{2\phi\over\sqrt{2}}+{2\sigma\over\sqrt{6}}}|F^{\prime}_{2}|^{2}
{-}{1\over12}{\epsilon^{\mu\nu\rho\sigma\gamma}}X_{\mu\nu\rho}      
{F^{\prime}_{\sigma\gamma}}
\eqn\onem
$$

Next we need to linearly combine the dilaton $\phi$ with the
KK scalar $\sigma$ so that ${\phi\over{\sqrt2}}
+{\sigma\over{\sqrt6}} = const\psi$. The constant is chosen so that
$U\epsilon$ SO(2) where
$$
U={1\over const}\pmatrix {{1\over{\sqrt2}} &{1\over{\sqrt6}} \cr
{-1\over{\sqrt6}} & {1\over{\sqrt2}} \cr}
\eqn\onen
$$
This requires the $const=\sqrt{2\over3}$, i.e.
$$
\pmatrix{\psi \cr \psi_{\perp} \cr} = \pmatrix{{\sqrt{3}\over{2}}
&{1\over2} \cr -{1\over{2}} & {\sqrt{3}\over{2}} \cr}
\pmatrix{\phi \cr \sigma \cr}
\eqn\oneo
$$
Truncating out the orthogonal combination ($\psi_{\perp} = 0$), we have

$\sigma = {\psi\over2} $ and $ \phi = {\sqrt{3}\over2}\psi $ .

\noindent
The resulting action is 
$$
\eqalign{
S_{5} = \int d^{5}x e\big\{ R_{5} -{1\over2}{|d\psi|}^{2} 
-{1\over4}e^{-{2\psi\over\sqrt{6}}} [ |F_{2}|^{2}+|G_{2}^{(1)}|^{2} 
+|{f_{2}}|^{2}+ |F^{\prime}_{2}|^{2} ] 
-{1\over4}e^{4\psi\over\sqrt{6}}|C_{2}|^{2} \cr +
4{\alpha}^{2}e^{2\psi\over\sqrt{6}} \big\} 
+ 
 {1\over2}\int C_{2} [G_{2}^{(1)}B_{1}^{(1)} +
Tr(F_{2}A_{1} -{i\alpha\over3}A_{1}A_{1}A_{1}) {+}
2F^{\prime}_{2}{\tilde A} ]}
\eqn\onep
$$
We still need to remove a vector to obtain a pure N=4 supergravity multiplet.
But we have the choice of truncating out any of the 2-form field
strengths $f_2$,${F_2}^{\prime}$,$G_2$,$F_2$,$C_2$ or any linear combination of
them. Since we ultimately want to reduce the model to 4-D and compare
with the 'half gauged' FS model [\e], we don't touch $F_2$ or $G_2$
as they already have the required Chern-Simons terms. Since $C_2$'s
kinetic term has a different dilatonic prefactor to those of the other
2-forms we also leave $C_2$ alone. This suggests that we should
linearly combine $f_2$ and ${F_2}^{\prime}$ and truncate the
orthogonal linear combination. This procedure, i.e. linearly combining
the dualisation of an antisymmetric tensor potential (which is a
remnant of the 3-form potential
$A_{\mu\nu\rho}$ from 7-D) with the KK vector, was actually
performed in 6-D in [\m] and will also have to be
performed in 4-D to obtain the 'half gauged' FS model.
This is done by another SO(2) rotation ,
$ F^{\prime}_{2} = {{G_{2}^{(2)} + H_{2}}\over\sqrt2}$ and 
$ {f_{2}} = {{G_{2}^{(2)} - H_{2}}\over\sqrt2}$ 
where $G_{2}^{(2)}=dB_{1}^{(2)}$ and $H_{2}=dh_{1}$. We can then
consistently truncate a vector, $H_2=0$.

After an integration by parts the resulting SU(2)$\times$U(1)$^2$
 gauged bosonic action in 5-D is : 
$$
  \eqalign{
S_{5} = \int d^{5}x e\big\{ R_{5} -{1\over2}{|d\psi|}^{2}
-{1\over4}e^{-{2\psi\over\sqrt{6}}} [ |F_{2}|^{2} + |G_{2}^{(1)}|^{2}
+ |G_{2}^{(2)}|^{2} ] 
-{1\over4}e^{4\psi\over\sqrt{6}} |C_{2}|^{2} \cr 
+ 4{\alpha}^{2} e^{2\psi\over\sqrt{6}} \big\} 
-{1\over2} \int C_{1} [ G_{2}^{(1)} G_{2}^{(1)}
+ G_{2}^{(2)} G_{2}^{(2)} + Tr(F_{2}F_{2}) ] }
\eqn\oneq
$$          
where $G_{2}^{(p)} = dB_{1}^{(p)} $ (p=1,2), $C_2=dC_1$ and  

\noindent
${F_{\mu\nu i}}^{j} = 2(\partial_{[\mu}{A_{\nu ] i}}^{j} +
i\alpha{A_{[\mu |i|}}^{k}{A_{\nu ] k}}^{j})$ (i,j=1,2).  

{\bf \chapter{\bf Dimensional Reduction To D=4}}

We proceed as in the previous case. In the following $\tilde A$ and
$\sigma$ are now new KK fields. Since they don't appear in (2.17) we
will use the symbols again. The ansatz for the fields are :
$$
{\hat e}^{\hat a}_{\hat m} = \pmatrix{ e^{\sigma\over2\sqrt3}e^{a}_{m}&0\cr
e^{-{\sigma\over\sqrt3}}\tilde A_{m} & e^{-{\sigma\over\sqrt3}}\cr }
\eqn\twoa
$$

Where hats refer to D=5.
$$
d{\hat S_{5}}^{2}=e^{\sigma\over\sqrt3}dS_{4}^{2}+e^{-{2\sigma\over\sqrt3}}
(dz+\tilde A)^{2}
\eqn\twob
$$
where $ \tilde A = \tilde A_{m}dx^{m}$ and $ f_{2} =d\tilde A.$
$$
{{{\hat A}}_{1 i}}^{\; \; \; j} = {A_{1 i}}^{j} + {A_{i}}^{j} dz
\eqn\twoc
$$
$$
{\hat C_{1}} = C_{1} + Bdz
\eqn\twod
$$
$$
{\hat B_{1}^{(p)}} = B_{1}^{(p)} + {\rho}^{(p)}dz
\eqn\twoe
$$
$$
\hat \psi (x^m,z) =  \psi (x^m)
\eqn\twof
$$
where (p)=1,2 and i,j=1,2

\noindent
The resulting action in 4-D is (after setting
${A_{i}}^{j}={\rho}^{(p)}=0$ )
$$
\eqalign{
S_{4} = \int d^{4}x e\big\{ R_{4}-{1\over2}{|d\sigma|}^{2}
-{1\over2}{|d\psi|}^{2}
-{1\over4}e^{-{3\sigma\over\sqrt3}}|{f_{2}}|^{2}
-{1\over4}e^{{4\psi\over\sqrt6}-{\sigma\over\sqrt3}}|C_{2}|^{2} \cr
-{1\over2}e^{{4\psi\over\sqrt6}+{2\sigma\over\sqrt3}}|dB|^{2} 
-{1\over4}e^{-{2\psi\over\sqrt6}-{\sigma\over\sqrt3}} [ |F_{2}|^{2}+|G_{2}^{(1)}|^{2}+|G_{2}^{(2)}|^{2} ] 
+ 4{\alpha}^{2}e^{{2\psi\over\sqrt6}+{\sigma\over\sqrt3}} \big\} \cr
-{1\over2} \int B [ G_{2}^{(1)} G_{2}^{(1)}
+ G_{2}^{(2)} G_{2}^{(2)} + Tr(F_{2}F_{2}) ] }
\eqn\twog
$$
where $f_2=d{\tilde A}_2$ , $C_2=dC_1-dB{\tilde A}$ ,
${G_2}^{(p)}=d{B_1}^{(p)}$, and ${F_{\mu\nu i}}^{j} = 2(\partial_{[\mu}{A_{\nu ] i}}^{j} +
i\alpha{A_{[\mu |i|}}^{k}{A_{\nu ] k}}^{j})$. 

 As above, we need to linearly combine $\psi$ with the
KK scalar $\sigma$. The procedure is the same as that
described for 5-D except 
$$
U= \pmatrix {{2\over{\sqrt6}} &{1\over{\sqrt3}} \cr
{-1\over{\sqrt3}} & {2\over{\sqrt6}} \cr} 
\eqn\twoh
$$
This leads to $\sigma = {1\over\sqrt3}\phi $ and $\psi = {\sqrt{2\over{3}}}\phi $.

In order to fully decouple an N=4 D=4 vector supermultiplet (1 vector,
6 scalars ) we need to dualise $C_{\mu}$ (the remnant in 4-D of the 7-D
field $A_{\mu\nu\rho}$) to $C^{\prime}_{\mu}$, linearly combine with
$\tilde A$ and truncate the orthogonal combination. This is suggested by
what has already been done in 6-D and 5-D. The procedure as
described above (with $a={+}{1\over4}$) replaces 
$$
L = -{1\over4}e^{{4\psi\over\sqrt6}-{\sigma\over\sqrt3}}|C_{2}|^{2}
\eqn\twoi
$$
with
$$
L_D = -{1\over4}e^{-{4\psi\over\sqrt6}+{\sigma\over\sqrt3}}|C^{\prime}_{2}|
{+}{1\over2}{\epsilon}^{\mu\nu\rho\sigma}{\partial}_{\mu}B{{\tilde
A}_{\nu}} {C^{\prime}}_{\rho\sigma}
\eqn\twoj
$$
where $C^{\prime}_{2} = d{C^{\prime}_{1}}$. Then $C^{\prime}_{1}$ is linearly
combined with ${\tilde A}$ using the same SO(2) matrix as in 5-D, i.e.
$$
\pmatrix{{f_{2}} \cr C^{\prime}_{2}} =
\pmatrix{ 1\over{\sqrt2} & {-1}\over{\sqrt2} \cr 1\over{\sqrt2} &
1\over{\sqrt2}}\pmatrix{ G^{(3)}_{2} \cr H_2}.
\eqn\twok
$$
After truncating the abelian field strength $H_2$, the contribution to the Chern-Simons term
from (3.10) is 
$$
- {1\over2} \int B G^{(3)}_{2} G^{(3)}_{2} .
\eqn\twol
$$

So then the N=4 vector supermultiplet decouples from the SU(2) gauged
N=4 supergravity multiplet and the resulting action is 
$$
\eqalign{
S_{4} = \int d^{4}x e\big\{ R_{4} -{1\over2}{|d\phi|}^{2} 
-{1\over2}e^{2\phi}{|dB|}^{2} + 4{\alpha}^{2}e^{\phi} \cr
 -{1\over4}e^{-{\phi}}
[ |F_{2}|^{2} + |G_{2}^{(1)}|^{2} +|G_{2}^{(2)}|^{2} + |G_{2}^{(3)}|^{2} ]
 \big\} \cr
-{1\over2}\int B [ Tr(F_{2}F_{2})
+G_{2}^{(1)}G_{2}^{(1)} + G_{2}^{(2)}G_{2}^{(2)} +
G_{2}^{(3)}G_{2}^{(3)} ] }
\eqn\twom
$$
where $G_{2}^{(p)} = dB_{1}^{(p)}$  (p=1,2,3) and ${F_{\mu\nu i}}^{j} = 2(\partial_{[\mu}{A_{\nu ] i}}^{j} +
i\alpha{A_{[\mu |i|}}^{k}{A_{\nu ] k}}^{j})$  (i,j=1,2). 

With the identification $e_{A} = {\sqrt{2}}\alpha$, we recognise this
action as the bosonic sector of the FS model [\e] with
one of the SU(2) coupling constants equal to zero.

The model contains both an abelian and a non-abelian sector. The
abelian sector has three U(1) field strengths $G_{2}^{(p)}$, whose
origins lie with the four form field strength in seven dimensions. The
non-abelian sector contains a triplet of SU(2) gauge fields with field
strengths ${F_{\mu\nu i}}^{j}$ coming straight down from the
corresponding SU(2) triplet in seven dimensions.

{\bf \chapter{\bf A Supersymmetric Electrovac In 7-D}}

So far the 7-D SU(2) gauged supergravity [\c] has been reduced from
seven to four dimensions on T${^3}$. In four dimensions the
reduction results in a version of the SU(2)$\times$SU(2) gauged N=4
supergravity of Freedman and Schwarz in which one of the gauge
coupling constants is zero.

In recent years there has been much interest in p-brane solutions of
supergravity theories. We will see later that these non-maximal SU(2) 
gauged supergravities posses ground state solutions which have the 
interpretation of a domain wall (D-2 brane) preserving a half of the 
supersymmetry. The domain wall in 7-D being the lift of the domain
wall solution of the FS theory.

Gibbons and Freedman [\f] however, had previously found a different type
of ground state solution of the 4-D SU(2)$_A$$\times$SU(2)$_B$ gauged
theory, namely a supersymmetric electrovac solution. Now that we have 
shown the $e_B=0$ version of the FS model to have a
seven dimensional origin, the existence of a supersymmetric electrovac
solution in four dimensions means that there must exist a similar
solution preserving a half of the supersymmetry in 7-D 
(and of course in 5 and 6-D) which is precisely the lift to 7-D of
this 4-D supersymmetric electrovac.

In this section therefore we will concentrate on constructing directly
in 7-D an electrovac solution of the N=2 SU(2) gauged supergravity
preserving a half of the supersymmetry. Setting $e_B$ to
zero, the Freedman/Gibbons electrovac (constant electric and magnetic fields)
preserves no supersymmetry if there are both electric and magnetic
fields present, and a half of the supersymmetry if it involves only
electric fields. The spacetime background being in the former case 
(AdS)$_{2}$$\times$S$^{2}$ and in the latter case
(AdS)$_{2}$$\times$${\R}^{2}$. 
The gauge fields of the non-abelian sector are zero, as is the pseudoscalar
B. The dilaton is constant and the constant electric and magnetic
fields arise from the abelian sector of the model. This indicates that
the solution we are seeking in 7-D only involves the fourth rank
antisymmetric tensor (the abelian sector) and the dilaton (which is
constant). Our strategy will be to find solutions of the Killing 
spinor equations.

We note at this point that there exist non-supersymmetric electrovacs
(involving magnetic fields) in 4-D which could also be lifted to 7-D
to give solutions. These solutions would be distinct from the 7-D
non-supersymmetric electrovacs found in [\n]. We shall be
concerned only with the supersymmetric electrovacs.

Consider first the background spacetime. One might have supposed it
would be (AdS)$_{2}$$\times$${\R}^{2}$$\times$T$^{3}$. 
However, the metric is not simply diagonal and this is crucial to 
finding the supersymmetric electrovac solution in 7-D. In [\n]
Quevedo used a diagonal metric ansatz and the resulting electrovacs
were non-supersymmetric. The reason that the metric contains off
diagonal terms is because the U(1) gauge potentials in 4-D responsible
for the constant electric fields of the Gibbons/Freedman solution are
linear combinations of vector fields arising from the four form in 7-D
and the KK vectors from the metric i.e. the KK vectors are 
not zero in the electrovac
solution. Thus there is a mixing between the vector fields from the
metric with those from the four form at each step of the reduction
from 7-D to 4-D

Having already performed the reduction we can now write the seven
dimensional metric in terms of the non-zero fields appearing in the
Gibbons/Freedman 4-D electrovac.

$$
dS_{7}^{2} = e^{{3\over{5}}{\phi}}dS_{4}^{2} + e^{-{{2\over{5}} {\phi}}}{(dZ_5 +
{{B_1}^{(3)}\over{\sqrt{2}}})^{2}} + e^{-{{2\over{5}}{\phi}}}{(dZ_6 +{{B_1}^{(2)}\over{\sqrt{2}}})^{2}} + e^{-{{2\over{5}}{\phi}}}{(dZ_7 +
{{B_1}^{(1)}\over{\sqrt{2}}})^{2}}  
\eqn\foura
$$
In the coordinates of [\f] $dS_{4}^{2}$, the line element of 
(AdS)$_2$$\times$${\R}^2$, is 
$$     
dS_{4}^{2} = {1\over{K{cos}^{2}\rho}} (-dt^2 + {d\rho}^2) + (dX^{2})^2 +
(dX^{3})^2 
\eqn\fourb
$$
where $K$ is the Gaussian curvature given by $K=2{e_A}E$, with $E$ a
constant to be identified as the electric field later. In order to
find expressions for ${B_1}^{(i)}$ we need to consider the 4-D
solution. The relevant part of the FS action is
$$
I_{FS} = \int d^{4}x \sqrt{-g} [R -{1\over2}(\partial\phi)^{2} +
{e^{-{\phi}}\over{4}} \Sigma \; |{G_2}^{(i)}|^{2} -2{e_{A}}^{2}e^{\phi}]
\eqn\fourc
$$
where the summation is over (i) from one to three. The $\phi$ equation of motion is
$$
{1\over{\sqrt{-g}}}{{\partial}_{\mu}}({\sqrt{-g}}{{\partial}^{\mu}}{\phi})
= {1\over{4}}e^{-{\phi}}{\Sigma}\; |{G_2}^{(i)}|^{2}
+2{e_{A}}^{2}e^{\phi} = {V_{eff}^{'}}(\phi)
\eqn\fourd
$$
For ${V_{eff}(\phi)}$ to have a minimum we require
$$
{\Sigma}\; {G_{\mu\nu}^{(i)}}{G^{\mu\nu}}^{(i)} = -8E^2
\eqn\foure
$$
then ${V_{eff}^{'}}(\phi) =0 \;$ when $e^{\phi}={E\over{e_{A}}}$ hence
$E$ is identified as the electric field.
Therefore for simplicity and without loss of generality we can choose

$ 
\; \; \; \; \; \; \; \; \; \; \; \; \; \; \; \; \; \; \; \; \; \; \;
{B_{1}^{(2)}}={B_{1}^{(3)}}=0 \; \; \; \; \; \; $and$ \; \; \; \; \;
\; 
{G_{\mu\nu}^{(1)}}{G^{\mu\nu}}^{(1)} = -8E^2 \; \; \; \; \; \; \; \; \;
\; \; \; \; \; \; \; \;  
\eqname\fourg $
A suitable potential is 
$$
B_{t}^{(1)} = {2Etan{\rho}\over{K}}
\eqn\fourh
$$
all other components vanishing (remember, in the reduction of
$B_{1}^{(1)}$ we set the scalars ${\rho}^{(1)}$ to zero).

Putting this all together with a relabelling X$^{4}=$Z$_{5}$,
X$^{5}=$Z$_{6}$ and X$^{6}=$Z$_{7}$ we have :
$$
\eqalign {
ds_{7}^{2} = (-{{e^{3\phi\over{5}}}\over{Kcos^{2}{\rho}}} + 
{e^{-{2\phi\over{5}}}}{{2E^{2}tan^{2}\rho}\over{K^{2}}} )dt^{2} + 
{{e^{3\phi\over{5}}}\over{Kcos^{2}{\rho}}}d{\rho^{2}} +
{e^{3\phi\over{5}}}(dX^{2})^{2} + {e^{3\phi\over{5}}}(dX^{3})^{2} \cr  
+ {e^{-{2\phi\over{5}}}}(dX^{4})^{2} + {e^{-{2\phi\over{5}}}}(dX^{5})^{2} +
 {e^{-{2\phi\over{5}}}}(dX^{6})^{2} + 
{e^{-{{2\phi}\over{5}}}}{{2{\sqrt{2}E}{tan{\rho}}}\over{K}} dtdX^{6} }
\eqn\fouri
$$
Choosing the vielbein frame as 
$$ 
\eqalign {e^{\underline{t}}=
{{e^{{3\phi}\over{10}}}\over{{\sqrt{k}}cos{\rho}}}dt \; \; \; \; \; \; \; \; \; \; \;
e^{\underline{\rho}}={{e^{{3\phi}\over{10}}}\over{{\sqrt{k}}cos{\rho}}}d{\rho}
\cr 
e^{\underline{2}}= {e^{{3\phi}\over{10}}}dX^{2} \; \; \; \; \; \; \; \; \; \; \; \; \; \; \; \; \; \;
\; \; 
\; e^{\underline{3}}= {e^{{3\phi}\over{10}}}dX^{3} \cr
e^{\underline{4}}= {e^{-{{\phi}\over{5}}}}dX^{4} \; \; \; \; \; \; \; \; \; \; \; \; \;
\; \; \; \; 
\; e^{\underline{5}}= {e^{-{{\phi}\over{5}}}}dX^{5} \cr
e^{\underline{6}}={e^{-{{\phi}\over{5}}}}{{\sqrt{2}Etan{\rho}}\over{K}}dt +
{e^{-{{\phi}\over{5}}}}dX^{6} \cr }
\eqn\fourj
$$
the only non-zero components of the spin connection are computed to be
$$
\eqalign {\omega_{t{\underline{t}}{\underline{\rho}}} = 
-tan(1-{{{E^{2}}\over{K}}{e^{-{\phi}}}}) \; \; \; \; \; \; \; \; \; \;
\; \; 
{\omega_{6{\underline{t}}{\underline{\rho}}} = 
{{E}\over{\sqrt{2}}}{e^{-{\phi}}}}  \cr
\omega_{\rho{\underline{t}}{\underline{6}}} = {{E}\over{\sqrt{2K}cos{\rho}}}
{e^{-{{\phi}\over{2}}}} {\; \; \; \; \; \; \; \;} 
\omega_{t{\underline{6}}{\underline{\rho}}} = {{E}\over{\sqrt{2K}cos{\rho}}}
{e^{-{{\phi}\over{2}}}} \cr}
\eqn\fourk
$$
where we have denoted 'flat space' Lorentz indices by underlining.

Next we need to understand which components of $F_{\mu\nu\rho\sigma}$
are non-vanishing. In the reduction from 7-D to 6-D $F_{4}$ reduced as
follows
$$
F_{\mu\nu\rho\sigma} = F_{\mu\nu\rho\sigma}^{'} + F_{\mu\nu\rho}dX^{6}
\eqn\fourl
$$ 
$F_{\mu\nu\rho\sigma}^{'}$ was then dualised to a two form which,
after linearly combining with the KK vector, became
$G_{\mu\nu}^{(1)}$. Hence $F_{\mu\nu\rho\sigma}^{'}$ is given by
$$
F_{\mu\nu\rho\sigma}^{'} = 
-{e\over{2\sqrt{2}}} {e^{-{{\psi}\over{\sqrt{2}} } } }
{\epsilon_{\mu\nu\rho\sigma\lambda\tau}}G^{\lambda\tau (1)} 
\eqn\fourm
$$
where $\psi$ and $G_{2}^{(1)}$ are 6-d fields and $\mu
=0,{\dots},5$. The further reduction of $\psi$ and $G_{2}^{(1)}$ to
4-D is straight forward.

In the reduction from 6-D to 5-D $F_{\mu\nu\rho}$ was reduced as 
$$
F_{\mu\nu\rho} = F_{\mu\nu\rho}^{'} + C_{\mu\nu}dX^{5} = 
F_{\mu\nu\rho 6}
\eqn\fourn
$$
$F_{\mu\nu\rho}^{'}$ was then dualised to a two form and linearly
combined with a KK vector so that 
$$
F_{\mu\nu\rho}^{'} = {e\over{2\sqrt{2}}} {e^{-{{{2}\over{\sqrt{6}}}{\psi}^{'}}}}
{\epsilon_{\mu\nu\rho\sigma\lambda}}G^{\sigma\lambda (2)} 
\eqn\fouro
$$
where ${\psi^{'}}$ and $G_{2}^{(2)}$ are 5-d fields and $\mu
=0,{\dots},4$. The further reduction of $G_{2}^{(2)}$ to
4-D is straight forward.

Similarly after the reduction from 5-D to 4-D $C_{\mu\nu}$ becomes
$$
C_{\mu\nu} = C_{\mu\nu}^{'} + {\partial}_{\mu} BdX^{4} = F_{\mu\nu 65}
\eqn\fourp
$$
with 
$$
C_{\mu\nu}^{'} = {e\over{2\sqrt{2}}}e^{-{\phi}}
{\epsilon_{\mu\nu\rho\sigma}}G^{\rho\sigma (3)} 
\; \; \; \; \; \mu = 0,{\dots},3
\eqn\fourq
$$

Now since $G_{2}^{(2)} = G_{2}^{(3)} = B =0$ in the 4-D electrovac and
since the only non-zero component of  $G_{2}^{(1)}$ is $G^{(1){\underline{0}}{\underline{1}}}$
this implies the only non-zero component of $F_{4}$ is 
$$
F_{{\underline{2}}{\underline{3}}{\underline{4}}{\underline{5}}}
\; \; \propto
\epsilon_{{\underline{0}}{\underline{1}}{\underline{2}}{\underline{3}}{\underline{4}}{\underline{5}}}G^{(1)\underline{0}\underline{1}}
\eqn\fourr
$$

We can now attempt to solve the 7-D Killing spinor equations. They are
$$
\eqalign{
\delta\lambda_{i} =
{1\over{2\sqrt{2}}}{\Gamma}^{M}D_{M}\Phi{\epsilon}_{i}- 
{{i{\sigma}_{0}}\over{2\sqrt{20}}}{\Gamma}^{MN}{F_{MNi}}^{j}{\epsilon}_{j}
+
{{{\sigma}_{0}}^{-2}\over{24\sqrt{20}}}{\Gamma}^{MNPQ}F_{MNPQ}{\epsilon}_{i}
\cr
+ {{{\alpha}{{\sigma}_{0}}^{-1}}\over{\sqrt{5}}}{\epsilon}_{i}}
$$ 
$$
\eqalign {\delta{\psi}_{Mi} = D_{M}{\epsilon}_{i} + {{i{\sigma}_{0}}\over{20}}
({\Gamma}_{M}^{NP} - 8{\delta}_{M}^{N}{\Gamma}^{P}){F_{NPi}}^{j}{\epsilon}_{j}+
{i{\alpha}\over{\sqrt{2}}}{A_{Mi}}^{j}{\epsilon}_{j}   \cr
{{{\sigma}_{0}^{-2}}\over{160}}({\Gamma}_{M}^{NPQR} - {8\over3}{\delta}_{M}^{N}
{\Gamma}^{PQR})F_{NPQR}{\epsilon}_{i} -
{{\alpha}{\sigma}_{0}^{-1}\over{5}} {\Gamma}_{M}{\epsilon}_{i} }
\eqn\fours
$$
where ${\sigma}_{0} = e^{-{{\Phi}\over{\sqrt{10}}}}$.

With ${F_{MNi}}^{j} =0$, $\Phi=$constant and using the lemma 
$$
{{\Gamma}_{M}}^{NPQR}={\Gamma}_{M}{\Gamma}^{NPQR}-{{\delta}_{M}}^{N}
{\Gamma}^{PQR}+{{\delta}_{M}}^{P}{\Gamma}^{QRN}-
{{\delta}_{M}}^{Q}{\Gamma}^{RNP}+{{\delta}_{M}}^{R}{\Gamma}^{NPQ}
\eqn\fourt
$$
they become
$$
\delta{\lambda}_{i}=
{{\Gamma}^{{\underline{2}}{\underline{3}}{\underline{4}}{\underline{5}}}F_{{\underline{2}}{\underline{3}}{\underline{4}}{\underline{5}}}}
{\epsilon}_{i}+ 2{\alpha}{\sigma}_{0}{\epsilon}_{i}
$$
$$
{\delta{\psi}_{Mi}}= {{\partial}_{M}}{\epsilon}_{i}+ {1\over4}
{\omega}_{M{\underline{a}}{\underline{b}}}
{\Gamma}^{{\underline{a}}{\underline{b}}}{\epsilon}_{i}+
{{{\sigma}_{0}^{-2}}\over{96}}[{\Gamma}_{M}{\Gamma}^{N}-4{\delta}_{M}^{N}]
{\Gamma}^{PQR}F_{NPQR}{\epsilon}_{i}
\eqn\fouru
$$
Since in 7-D the product of all the gamma matrices is 1, a useful
expression is 
$$
{{\Gamma}^{{\underline{2}}{\underline{3}}{\underline{4}}{\underline{5}}}}
=
-{{\Gamma}^{\underline{\rho}}}{{\Gamma}^{\underline{t}}}{{\Gamma}^{\underline{6}}}
\eqn\fourv
$$

The 7-D dilaton is related to the 4-D dilaton by a scaling,
$\Phi={\sqrt{2\over{5}}}\phi$ hence
${\sigma}_{0}=e^{-{{\phi}\over{5}}}$.
Therefore for $\delta{\lambda}_{i}=0$,  $F_{{\underline{2}}{\underline{3}}{\underline{4}}{\underline{5}}}
$ must satisfy
$$
{{\Gamma}^{{\underline{2}}{\underline{3}}{\underline{4}}{\underline{5}}}F_{{\underline{2}}{\underline{3}}{\underline{4}}{\underline{5}}}
{\epsilon}_{i}= -2{\alpha}e^{-{{\phi}\over{5}}}
{\epsilon}_{i}}
\eqn\fourw
$$
Consider the ${\delta{\psi}_{ti}}$ equation. Using (4.10) we have
$$
\eqalign{
\delta{\psi}_{ti}={\partial}_{t}{\epsilon}_{i}-{{tan{\rho}}\over{2}}
(1-{e^{-{\phi}}{{{E^{2}}}\over{K}}}){\Gamma}^{{\underline{t}}{\underline{\rho}}}
{\epsilon}_{i}+{e^{-{{\phi}\over{2}}}}{E\over{{2\sqrt{2K}}cos{\rho}}}
{\Gamma}^{{\underline{6}}{\underline{\rho}}}{\epsilon}_{i} \cr
{{e^{2{\phi}\over{5}}}\over{4}}(e_{t}^{\underline{t}}{\Gamma}_{\underline{t}}
+e_{t}^{\underline{6}}{\Gamma}_{\underline{6}){\Gamma}^{{\underline{2}}{\underline{3}}{\underline{4}}{\underline{5}}}F_{{\underline{2}}{\underline{3}}{\underline{4}}{\underline{5}}}{\epsilon}_{i}}}
\eqn\fourx
$$
Substituting the expressions for the vielbein components (4.9), the equation
becomes
$$
\eqalign{
\delta{\psi}_{ti}={\partial}_{t}{\epsilon}_{i}-{{tan{\rho}}\over{2}}
(1-{{e^{-{\phi}}}{{{E^{2}}\over{K}}}}){\Gamma}^{{\underline{t}}{\underline{\rho}}}
{\epsilon}_{i}+{e^{-{{\phi}\over{2}}}}{E\over{{2\sqrt{2K}}cos{\rho}}}
{\Gamma}^{{\underline{6}}{\underline{\rho}}}{\epsilon}_{i} \cr
-{e^{{7{\phi}\over{10}}}}{{{\Gamma}^{\underline{t}}}\over{{4\sqrt{K}}cos{\rho}}}
{{\Gamma}^{{\underline{2}}{\underline{3}}{\underline{4}}{\underline{5}}}}F_{{\underline{2}}{\underline{3}}{\underline{4}}{\underline{5}}}{\epsilon}_{i}
+{e^{{{\phi}\over{5}}}}{{Etan{\rho}}\over{{2\sqrt{2K}}}}{\Gamma}^{\underline{6}}
{{\Gamma}^{{\underline{2}}{\underline{3}}{\underline{4}}{\underline{5}}}F_{{\underline{2}}{\underline{3}}{\underline{4}}{\underline{5}}}{\epsilon}_{i}}}
\eqn\foury
$$
Using (4.21) in the last term, (4.22) in the fourth term and both
(4.21) and (4.22) in the third term we have
$$
\eqalign {
\delta{\psi}_{ti}={\partial}_{t}{\epsilon}_{i} +
{{{\Gamma}^{\underline{t}}}\over{2cos{\rho}}} \big [
-{e^{-{{3{\phi}}\over{10}}}}{E\over{{2\sqrt{2K}}\alpha}}F_{{\underline{2}}
{\underline{3}}{\underline{4}}{\underline{5}}}+
e^{-{{\phi}\over{2}}}{{\alpha}\over{\sqrt{K}}}\big ] {\epsilon}_{i} \cr
+{tan{\rho}\over{2}} \big [-1
+{e^{{-{\phi}}}}{{E^{2}}\over{K}}+
{e^{{{\phi}\over{5}}}}{E\over{{\sqrt{2}K}}}F_{{\underline{2}}{\underline{3}}{\underline{4}}
{\underline{5}}}\big ]{\Gamma}^{{\underline{t}}{\underline{\rho}}}
{\epsilon}_{i}}
\eqn\fourz
$$
Next consider $\delta{\psi}_{\rho i}$. Using (4.10) we have
$$
\delta{\psi}_{\rho i} = {\partial}_{\rho}{\epsilon}_{i} + 
e^{-{{\phi}\over{2}}}{E\over{2\sqrt{2K}cos{\rho}}}{\Gamma^{{\underline{t}}{\underline{6}}}}{\epsilon}_{i} + {{e^{2\phi\over{5}}}\over{4}}e_{\rho}^{\underline{\rho}}{\Gamma_{\underline{\rho}}}{{\Gamma}^{{\underline{2}}{\underline{3}}{\underline{4}}{\underline{5}}}F_{{\underline{2}}{\underline{3}}{\underline{4}}{\underline{5}}}{\epsilon}_{i}}
\eqn\fouraa
$$
Using the expression for $e_{\rho}^{\underline{\rho}}$ and equations
(4.21) and (4.22) we get
$$
\delta{\psi}_{\rho i} = {\partial}_{\rho}{\epsilon}_{i}
- {{\Gamma^{\underline{\rho}}}\over{2cos{\rho}}} \big [ 
e^{{\phi}\over{2}}{{\alpha}\over{\sqrt{K}}} - e^{-{{3\phi}\over{10}}}
{E\over{2\sqrt{2K}\alpha}}F_{{\underline{2}}{\underline{3}}{\underline{4}}{\underline{5}}}
\big ] {\epsilon}_{i}
\eqn\fourbb
$$
We will return to $\delta{\psi}_{\rho i}$ and $\delta{\psi}_{ti}$ later.
Next consider $\delta{\psi}_{2i}$, $\delta{\psi}_{3i}$,
$\delta{\psi}_{4i}$, and $\delta{\psi}_{5i}$. Assuming ${\epsilon}_{i}$ 
depends only on $\rho$ and $t$ we have
$$
\delta{\psi}_{2i} = {{e^{{2\phi}\over{5}}}\over{96}}\big [
4!{\Gamma_{2}}{{\Gamma}^{{\underline{2}}{\underline{3}}{\underline{4}}{\underline{5}}}F_{{\underline{2}}{\underline{3}}{\underline{4}}{\underline{5}}}}
-4{\delta_{2}^{\underline{2}}}{{\Gamma}^{{\underline{3}}{\underline{4}}{\underline{5}}}F_{{\underline{2}}{\underline{3}}{\underline{4}}{\underline{5}}}}(3!)
\big ] = 0
\eqn\fourcc
$$
The expressions for  $\delta{\psi}_{3i}$,
$\delta{\psi}_{4i}$ and $\delta{\psi}_{5i}$ vanish similarly.
Finally consider $\delta{\psi}_{6i}$.
$$
\delta{\psi}_{6i}=
e^{-\phi}{E\over{2\sqrt{2}}}{\Gamma^{{\underline{t}}{\underline{\rho}}}}{\epsilon}_{i}+{e^{{2\phi}\over{5}}\over{4}}e_{6}^{\underline{6}}{\Gamma_{\underline{6}}}
{{\Gamma}^{{\underline{2}}{\underline{3}}{\underline{4}}{\underline{5}}}F_{{\underline{2}}{\underline{3}}{\underline{4}}{\underline{5}}}{\epsilon}_{i}}
\eqn\fourdd
$$
again, substituting the expression for $e_{6}^{\underline{6}}$ and the
equations (4.21),(4.22) we have,
$$
\delta{\psi}_{6i}= \big [ -e^{-{{4\phi}\over{5}}}{E\over{4\sqrt{2\alpha}}}
F_{{\underline{2}}{\underline{3}}{\underline{4}}{\underline{5}}} -
{{\alpha}\over2} \big ]{\Gamma^{\underline{6}}}{\epsilon}_{i}
\eqn\fouree
$$
For $\delta{\psi}_{6i}=0$ we require 
$$
F_{{\underline{2}}{\underline{3}}{\underline{4}}{\underline{5}}}= 
-{e^{{4\phi}\over{5}}}{{{2\sqrt{2}}{\alpha}^{2}}\over{E}}
\eqn\fourff
$$
Using this result in (4.22) it becomes
$$
({{\Gamma}^{{\underline{2}}{\underline{3}}{\underline{4}}{\underline{5}}}}
- e^{-\phi}{E\over{{\sqrt{2}}{\alpha}}}){\epsilon}_{i}=0
\eqn\fourgg
$$
Now we know 
$$
e^{\phi}= {E\over{e_{A}}} \; \; \; \; \; \; \; \; \; {e_{A}}={\sqrt{2}{\alpha}}
\eqn\fourhh 
$$
hence $\epsilon_{i}$ must be of the form
$$
\epsilon_{i}={1\over{2}}({{\Gamma}^{{\underline{2}}{\underline{3}}{\underline{4}}{\underline{5}}}}
+1){\xi}_{i}
\eqn\fourii
$$
where ${\xi}_{i}$ is a constant spinor.

Considering again the $\delta{\psi}_{ti}$ and $\delta{\psi}_{\rho i}$
equations and using (4.31), (4.33) and $K=2{e_{A}}E$, they become :
$$
\delta{\psi}_{ti}={\partial}_{t}{\epsilon}_{i} +
{{\Gamma^{\underline{t}}}\over{2cos{\rho}}}{\epsilon_{i}} -
{tan{\rho}\over{2}}{\Gamma^{{\underline{t}}{\underline{\rho}}}}{\epsilon}_{i}
\eqn\fourjj
$$
$$
\delta{\psi}_{\rho i}={\partial}_{\rho}{\epsilon}_{i}
- {{\Gamma^{\underline{\rho}}}\over{2cos{\rho}}}{\epsilon_{i}}
\eqn\fourkk
$$
By comparison with the Freedman/Gibbons electrovac in 4-D we see that
if $\epsilon_{i}$ is of the form 
$$
\epsilon_{i} = S(t,{\rho}){1\over{2}}(1+
{{\Gamma}^{{\underline{2}}{\underline{3}}{\underline{4}}{\underline{5}}}}){\xi}_{i}
\eqn\fourll
$$
with ${\xi}_{i}$ a constant spinor and the function $S(t,{\rho})$
given by
$$
S(t,{\rho})= {1\over{{(cos{\rho})}^{1\over{2}}}} \big [
cos{\rho\over{2}}+ {\Gamma^{\underline{\rho}}}sin{\rho\over{2}} \big ]
\big [ cos{t\over{2}} - {\Gamma^{\underline{t}}}sin{t\over{2}} \big ]
\eqn\fourmm
$$
then $\delta{\psi}_{ti} = \delta{\psi}_{\rho i} = 0 $ and a half of the
supersymmetry remains unbroken.

To summarise the 1/2 supersymmetric 7-D electrovac solution we
choose $E=e_{A}$ so that $\phi = 0 $. Then the metric and four form
field strength take the forms:

$$
F_{2345} = -24{\sqrt{2}}e_{A}
\eqn\fournn
$$
$$
d{S_7}^2 = {1\over{2{e_{A}}^2{cos{\rho}}^{2}}} \big [ -dt^2 +
d{\rho}^2 \big ] + \big [ dX^6 + {tan{\rho}\over{\sqrt{2}e_{A}}}dt
\big ]^2 + dS^{2}({\E}^4).
\eqn\fouroo
$$
where ${\E}^4$ is a 4-d Euclidean space containing the 2,3,4 and 5
spacial directions and all other fields and components are vanishing.
We have shown that this metric is (ADS)${_3}\times{{\E}^4}$.

{\bf \chapter{\bf Reduction Of D=7 Domain Wall}}
In this section, we turn to a consideration of the (D-2)-brane
solutions that arise in gauged supergravities.

As was shown in [\h,\i], supergravities with bosonic sectors of the form 
$$
L =e[R-{1\over2}(\partial\phi)^{2}+{{m^{2}}\over2}e^{-a\phi} ]
\eqn\threea
$$
have domain wall solutions of the form
$$
ds^{2}= H^{4\over{{\Delta}(D-2)}}dx^{\mu}dx^{\nu}{\eta_{\mu\nu}}
+H^{4(D-1)\over{{\Delta}(D-2)}} dy^{2}
\eqn\threeb
$$
$$
e^{\phi} = H^{2a\over\Delta}
\eqn\threec
$$
where D is the spacetime dimension and $\mu,\nu = 1,..,D-1$. $H$ is a 
harmonic function on the 1-dimensional transverse space with
coordinate $y$, of the general form $H=c \pm My$, where $c$ is an
arbitrary constant, and $M={1\over2}m{\sqrt {-\Delta}}$. See [\i] for
details of the spacetime structure of $\Delta = -2$ domain walls.

\noindent

D=7 N=2 SU(2) gauged supergravity has the 5-brane solution (in the
Einstein frame) [\h] :  
$$
ds_{7}^{2}=H^{-{2\over5}}dx^{\mu}dx^{\nu}{\eta}_{\mu\nu} +
H^{-{12\over5}}dy^{2}
\eqn\threed
$$
$$
e^{\phi{_7}} = H^{\sqrt{2\over5}}
\eqn\threeda
$$
where $\mu ,\nu =0,..,5$,
for which $\Delta = -2$. Notice that unlike the domain walls of massive
supergravities [\k] where $\Delta$ is always positive, and equal to 4 when only
one dilaton is involved in the solution, the value of $\Delta$ for
these gauged supergravities is negative. This implies that these gauged
theories cannot be obtained by ordinary KK or Scherk-Schwarz
reductions on T$^n$ of eleven dimensional supergravity or massive
IIA supergravity [\k,\o,\b].

We now double dimensionally reduce this 5-brane to obtain a membrane
solution of the FS model in 4-D. Since $\Delta$ is
unchanged under dimensional reduction it also has $\Delta =-2$. This
implies that a=1 which is indeed the case for the FS 
model, demonstrating the consistency of the reduction.

The ansatz for the reduction from D=7 to D=6 is
$$
d{\hat S_{7}^{2}}=e^{\sigma\over\sqrt10}dS_{6}^{2}+
e^{-{4\sigma\over\sqrt10}}dz^{2}
\eqn\threee
$$
All fields are assumed to be independent of the compactification 
coordinate z. Therefore z must be identified with a world volume
coordinate $x_{5}$ implying
$$
e^{-{4\sigma\over\sqrt10}}=H^{-{2\over5}}
\eqn\threef
$$
This then leads very simply to the D=6 4-brane [\h] :
$$
ds_{6}^{2}=H^{-{1\over2}}dx^{\mu}dx^{\nu}{\eta}_{\mu\nu} +
H^{-{5\over2}}dy^{2}
\eqn\threeg
$$
$$
e^{\phi{_6}} = H^{\sqrt{1\over2}}
\eqn\threega
$$
where $\mu ,\nu =0,..,4$.

\noindent
In the same way, using the ansatz (2.3) and (3.2)  (with $\tilde A=0$)
one can double dimensionally reduce this D=6 4-brane to a D=5 3-brane :
$$
ds_{5}^{2}=H^{-{2\over3}}dx^{\mu}dx^{\nu}{\eta}_{\mu\nu} +
H^{-{8\over3}}dy^{2}
\eqn\threeh
$$
$$
e^{\phi{_5}} = H^{\sqrt{2\over3}}
\eqn\threeha
$$
where $\mu ,\nu =0,..,3$

\noindent
and a D=4 2-brane :
$$
ds_{4}^{2}=H^{-{1}}dx^{\mu}dx^{\nu}{\eta}_{\mu\nu} +
H^{-{3}}dy^{2}
\eqn\threei
$$
$$
e^{\phi{_4}} = H
\eqn\threeia
$$
where $\mu ,\nu =0,..,2$.

{\bf \chapter{\bf Supersymmetric Domain Walls Of The Freedman/Schwarz Model}}

In this section we consider the supersymmetry properties of the
membrane solution of the FS model. As in [\e] we let $e_A$ and $e_B$
denote the two gauge coupling constants and we define
$$
e^2= e_A^2 + e_B^2
\eqn\foura
$$
The potential of this model has the form $V= -2{e^2}e^{\phi}$, so
$\Delta=-2$.
This model therefore has a domain wall solution of the type given in
\threei ,\threeia.
i.e.
$$
\eqalign{
ds^2 &= H^{-1}(y) dx^{\mu}dx^{\nu}{\eta_{\mu\nu}} +H^{-3}(y) dy^2\cr
e^\phi &= H(y) } 
\eqn\fourb
$$
where $H=m|y|+ constant$ and $m^2 = 2e^2 $.
Denoting `flat space' Lorentz indices by
underlining, the only non-zero components of the spin connection 
$\omega_{\mu\underline{a}\underline{b}}$ are
$$
\omega_{t{\underline{t}}{\underline{y}}} =
-\omega_{x{\underline{x}}{\underline{y}}}
=-\omega_{z{\underline{z}}{\underline{y}}} = {m\over2}
\eqn\fourc
$$
To check for supersymmetry, all that is needed here are the supersymmetry
transformations of the four Majorana gravitini $\psi$ and spin ${1\over2}$ fields $\chi$ in a bosonic background for which all
Yang-Mills field strengths vanish. In our `mostly plus' metric convention the
D=4 Dirac matrices can be assumed to be real. We define $\gamma_5$ to be the
product of all four Dirac matrices satisfying $\gamma_5^2 =-1$. In these
conventions the required supersymmetry transformations of the
FS model are 
$$
\eqalign{
\delta\psi_\mu &= 2\big[D_\mu \epsilon - {\sqrt{2}\over4}e^{\phi\over2} 
(e_A -\gamma_5 e_B) \Gamma_\mu \epsilon \big] \cr
\delta\chi &= {1\over\sqrt{2}}\big[ \partial_\mu \phi \Gamma^\mu +
{\sqrt{2}} e^{\phi\over2}(e_A +\gamma_5 e_B)\big]\epsilon }
\eqn\fourd
$$
The amount of supersymmetry preserved by the domain wall solution is the
number of independent solutions for $\epsilon$ of the conditions
$\delta\psi_\mu=0$ and $\delta\chi=0$ in the domain wall
background. Using $\epsilon = H^{-{1\over4}}(y)\epsilon_0 =
e^{{A(y)}\over2}\epsilon_0$, these conditions are
$$
\eqalign{
\delta\psi_\mu &= {-{1\over 2}}e^{A\over2}\big[-mH(y)^{1\over2} 
\Gamma^{\underline{y}} +{\sqrt{2}}e^{\phi\over2}(e_A -\gamma_5e_B)\big]
\Gamma_\mu \epsilon_0 =0 \qquad{\mu =x,z} \cr
\delta\psi_t &= {-{1\over 2}e^{A\over2}}\big[mH(y)^{1\over2} 
\Gamma^{\underline{y}} +{\sqrt{2}}e^{\phi\over2}(e_A -\gamma_5e_B)\big]
\Gamma_t \epsilon_0 =0 \cr
\delta\psi_y &= {e^{A\over2}}\big[ \partial_y A(y) -{\sqrt{2}\over2} 
e^{\phi\over2}(e_A -\gamma_5 e_B)\Gamma_y \big]\epsilon_0 =0 \cr 
\delta\chi &= {1\over\sqrt{2}}{ e^{A\over2}}\big[ \partial_y \phi+{\sqrt{2}}
 e^{\phi\over2}(e_A +\gamma_5 e_B)\Gamma_y\big]\Gamma^y\epsilon_0 =0\ . }
\eqn\foure
$$
They become
$$
\eqalign{
\delta\psi_\mu &= {-{1\over2}}e^{A\over2}\Gamma^{\underline{y}}(e_{A}+
\gamma_{5}e_{B})\big[\pm {m\over{e}}H(y)^{1\over2}+{\sqrt{2}}e^{\phi\over2}
\big]\Gamma_{\underline{y}}\Gamma_{\mu}\epsilon_0 =0 \qquad{\mu =x,z,t} \cr
\delta\psi_y &= {e^{A\over2}}\big[\partial_y A(y)\mp {\sqrt{2}\over2}
e{e^{\phi\over2}}{H(y)}^{-{3\over2}}\big]\epsilon_0 =0 \cr
\delta\chi &={1\over\sqrt{2}}{e^{A\over2}}\Gamma^y\big[\partial_y\phi\pm 
{\sqrt{2}}ee^{\phi\over2}{H(y)}^{-{3\over2}}\big]\epsilon_0 =0 \ } 
\eqn\fourf
$$
provided the constant spinor $\epsilon_0$ satisfies
$$
e^{-1}(e_A -\gamma_5 e_B)\Gamma_{\underline {y}}\epsilon_0 = \pm \epsilon_0\ .
\eqn\fourg
$$ 
The conditions on $\phi(y)$ and $A(y)$ for preservation of supersymmetry are
the same as demanded by the domain wall solution. The condition (6.7)
implies that half the supersymmetry is preserved.

In the case $e_B =0$ the above supersymmetry transformation
laws are much simplified and the domain wall preserves a half of the 
supersymmetry provided the constant spinor $\epsilon_0$ satisfies
$$
\Gamma_{\underline {y}}\epsilon_0 = \pm \epsilon_0\ .
\eqn\fourg
$$ 
It was shown in [\h] that the killing spinors of the 7-D 5-brane and
6-D 4-brane only depend on the transverse coordinates but are
independent of the compactification coordinate. Therefore in reducing
from 6-D to 5-D, there remains the same number of killing spinors,
hence the 5-D 3-brane also preserves half of the supersymmetry.

We note that unlike the electrovac groundstates where a gauge field is
non-vanishing and the dilaton is zero, domain walls involve no
gauge fields and the dilaton is non-vanishing.

{\bf \chapter {\bf Conclusion}}

Certain gauged supergravity theories contain dilatonic
potentials of the form $e^{-a\phi}$ and hence possess domain wall
solutions. Despite the absence of an $S^1\times M_{D-1}$ ground state
these theories can be consistently reduced [\i] to yield other gauged
supergravities in lower dimensions.

The SU(2)$\times$SU(2) gauged Freedman Schwarz model [\e] has
previously been identified as part of the effective D=4 field theory
for the heterotic string in an $S^3\times S^3$ vacuum [\p]. The `no-go'
theorem of Gibbons, Freedman and West [\q] is avoided due to that fact
that the D=4 dilaton is not presumed to be constant. However, at the
supergravity level it is not known which theory upon compactification
yields the FS model. In this paper we have
learnt that the SU(2)$\times$U(1)$^3$ gauged version can be obtained
by dimensional reduction of an SU(2) gauged supergravity in 7-D
[\c]. This connection then implies the existence of a supersymmetric
electrovac in 7-D which we have found.  Currently there is no known
link between this 7-D gauged non-maximal supergravity and 11-D
supergravity. However, dimensional reduction of 11-D supergravity on
$S^{4}$ yields an SO(5) gauged 7-D supergravity [\r]. It is unclear as yet
whether the 7-D non-maximal theory is a truncation of this SO(5)
gauged model. If it is or if it is linked to D=11 via compactification
on a different space, then these supersymmetric electrovac ground
states and domain walls would be interesting new solutions of M-theory
and may have an interpretation in terms of branes or intersecting
configurations of branes.

In performing the reduction we have obtained the bosonic sector of an
SU(2)$\times$U(1)$^2$ 5-D gauged theory which presumably has a
supersymmetric extension. It is interesting to contrast this D=5 model
with the D=5 N=4 SU(2)$\times$U(1) gauged supergravity of Romans
[\s]. The bosonic sector of the D=5 model presented in section 2
includes a pair of vectors, $B_{\mu}^{(i)}$, which form a doublet
under a global SO(2) symmetry. The bosonic sector of the D=5 model of
Romans differs from this in that the doublet of vectors is replaced by
a doublet of second rank antisymmetric tensor potentials allowing the
SO(2) (hence SU(2)$\times$U(1)) symmetry to be gauged. This results in
an additional term in the potential proportional to the SO(2) coupling
constant $g_1$. The gauging is effected using `odd dimensional self
duality' (ODSD) [\t] but the ungauged limit can be recovered by eliminating
one of the second rank antisymmetric tensors via its ODSD
equation. This leads to a Lagrangian of the usual form for a single
massive second rank antisymmetric tensor from which the ungauged limit
can be obtained. This is similar to the situation arising in D=7
where the SO(5) gauged model of [\r], containing a 5-plet of massive,
self dual, third rank antisymmetric potentials, is understood in
principle to be related, using ODSD, to the
ungauged maximal D=7 supergravity of [\u] which contains a 5-plet of
second rank potentials. However, in this case the ungauged limit is
not recoverable.

It is also interesting to compare the D=5
supersymmetric magnetovac of [\s] with the lift to D=5 of the 1/2
supersymmetric electrovac of the `half gauged' FS theory. This will be
referred to as the D=5 Gibbons Freedman (GF) magnetovac. In the
electrovac of the `half gauged' FS model, the non-abelian sector and
the scalars were zero. The electric field was due to one of the three
remaining U(1) potentials non-vanishing. It preserved a 1/2 of the
supersymmetry and the background was ADS$_2{\times}{\R}^2$. The
relevant part of the action was,

$$
S_{4} = \int d^{4}x e\big\{ R_{4} + 4{\alpha}^{2} 
 -{1\over4}[ |G_{2}^{(1)}|^{2} +|G_{2}^{(2)}|^{2} + |G_{2}^{(3)}|^{2} ]
 \big\}. 
$$
Without loss of generality one could choose $G_{2}^{(3)} {\not=}0$ and 
$G_{2}^{(1)}=G_{2}^{(2)}=0$. Being an electrovac, the non-zero
components of $G_{2}^{(3)}$ were $G_{01}^{(3)}$. But we know that in
lifting this electrovac up to D=5 $G_{2}^{(3)}$ came from
$C_{2}$. Hence, as a dualisation was involved, $C_{2}$ has only
space-space components and so the lift of this supersymmetric
electrovac is in fact a supersymmetric magnetovac. By the same token,
lifting up to six would result in a supersymmetric `electrovac' and in
D=7 we would get a supersymmetric `magnetovac'. As the only non-zero
component of $F_4$ supporting the D=7 `electrovac' of section 4 was
$F_{2345}$, perhaps this solution should have been called a `magnetovac'.
In the 5-D `magnetovac' of [\s] the second rank antisymmetric tensor
potentials are zero but the SU(2) and U(1) gauge fields are
non-vanishing. Hence, apart form the extra term in the scalar
potential proportional to $g_1$, the sectors of the model probed by
this solution are the same as the sectors of the D=5 model (2.17)
probed by the D=5 GF magnetovac in which $G_{2}^{(1)} =
G_{2}^{(2)}=0$. However, these two D=5 magnetovacs cannot be
identified as in the latter solution, the SU(2) gauge fields are
zero. One might suppose that a limit can be taken in which the SU(2)
gauge fields of Romans' magnetovac can be turned off. This is only
possible if $g_1$ can be sent to zero. However, the field equations
for the second rank potential are identically satisfied by choosing
these fields to vanish. In the remaining field equations $g_1$ can be 
sent to zero allowing the identification with the corresponding
remaining field equations of (2.17) and of the supersymmetric
magnetovac solutions to be made.

We have also shown that the 5-brane of the 7-D gauged theory [\h] can
be double dimensionally reduced to yield a domain wall in 4-D which is
a solution of the full SU(2)$\times$SU(2) gauged model preserving a half of
the supersymmetry. 

{\sl Acknowledgements.}
The author would like to acknowledge Paul Townsend 
for many valuable and ongoing discussions. He would also like to thank
his parents for continued support and Ronaldo Luiz Nazario De Lima 
who continues to inspire.

\refout
\end